\documentstyle[prd,aps]{revtex}
\tighten
\draft
\begin{document} 
\bibliographystyle{prsty}

 \mathchardef\ddash="705C

\preprint{TOKAI-HEP/TH-9901}
\title{
Dynamically Favored Chiral Symmetry Breakings\\
in Supersymmetric Quantum Chromodynamics
\footnote{\it To be published in Prog. Theor. Phys. {\bf 101} (April, 1999).}
}

\author{Yasuharu Honda$^{a}$
\footnote{E-mail:8jspd005@keyaki.cc.u-tokai.ac.jp}
and Masaki Yasu${\grave {\rm e}}^{a,b}$
\footnote{E-mail:yasue@keyaki.cc.u-tokai.ac.jp}
}

\address{\vspace{5mm}$^{a}${\sl Department of Physics, Tokai University}\\
{\sl 1117 KitaKaname, Hiratsuka, Kanagawa 259-1292, Japan}}
\address{\vspace{2mm}$^{b}${\sl Department of Natural Science\\School of Marine
Science and Technology, Tokai University}\\
{\sl 3-20-1 Orido, Shimizu, Shizuoka 424-8610, Japan}}
\date{TOKAI-HEP/TH-9901, February, 1999}
\maketitle

\begin{abstract}
Using an effective superpotential in supersymmetric quantum 
chromodynamics (SQCD) with $N_f$ flavors and $N_c$ 
colors of quarks for $N_f\geq N_c$+2, the influence of soft supersymmetry 
(SUSY) breakings is examined to clarify dynamics of chiral symmetry breakings near the 
SUSY limit. In the case that SQCD triggers spontaneous chiral symmetry breakings, 
it is possible to show that our superpotential dynamically 
favors the successive formation of condensates, leaving either $SU(N_f-N_c)$ or 
$SU(N_f-N_c+1)$ unbroken as a chiral nonabelian symmetry.
\end{abstract}


\section{\bf Introduction}
It has been argued that $N$=1 supersymmetric quantum chromodynamics 
(SQCD) with $N_f$ flavors and $N_c$ colors of massless quarks exhibits 
chiral $SU(N_f)$ $\times$ $U(1)$ symmetry which, in the confining 
phase, will be spontaneously broken for $N_f$ $\leq$ $N_c$ 
and will remain unbroken for $N_f$ $\geq$ $N_c$+1.  The existence of unbroken 
chiral symmetries are dynamically supported by non-perturbative 
superpotential\cite{Seiberg,EearlySeiberg} and 
algebraically by the t' Hooft anomaly-matching 
conditions\cite{tHooft}.  However, the suggested dynamics of SQCD with $N_f$ 
$\geq$ $N_c$+2 is based on a plausible extrapolation from those respecting 
the marvelous duality in $N$=2 supersymmetric (SUSY) gauge thoeries\cite{SeibergWitten}.  
To utilize this duality in SQCD, $\ddash$magnetic" quarks are set to contribute 
in the physics of SQCD.
The dynamics of $\ddash$magnetic" quarks is not derived, but arranged 
by our convenience so as to fulfill 
the algebraic requirement from the anomaly-matching conditions. 
Once such $\ddash$magnetic" quarks are admitted to participate in SQCD, 
every other physical consequences correctly follows and a massive 
number of consistency checks are realized\cite{Review}. 

However, what is the $\ddash$magnetic" quark?  The answer can be found in SQCD 
embedded in a softly broken $N$=2 SQCD\cite{CoulombPhase,BrokenN_2},
where the scale invariance of the $N$=2 theory plays a crucial role. 
The $\ddash$magnetic" quarks can be identified with those appearing in the 
$N$=2 SQCD.   Such a description of the $N$=1 duality in terms of the $N$=2 
duality is expected to be consistent with the correct physics in SQCD with 
3$N_c$/2 $<$ $N_f$ $<$ 3$N_c$, where the phase is characterized as an interacting 
Coulomb phase\cite{Seiberg}. On the other hand, in SQCD with 
$N_f$ $\leq$ 3$N_c$/2, where 
confinement will take place, no direct convincing derivation 
of the desired physics of $\ddash$magnetic" quarks has been made. It is thus 
reasonable to start seeking possibilities other than SQCD with $\ddash$magnetic" 
quarks, in other words, the possibility of  spontaneously 
broken chiral $SU(N_f)$ symmetry\cite{GapEquation}.

In SQCD with $N_f$ $\geq$ $N_c$+2, low-energy chiral symmetry 
of $\ddash$electric" quarks cannot sit on the origin of moduli space, where no vacuum 
expectation values are generated, since no set of composite superfields that 
perfectly satisfies the anomaly-matching conditions has been found.  
To maintain full chiral symmetry as "quantum" chiral symmetry, it has been 
conjectured\cite{Seiberg} that the theory experiences phase transition to 
its $\ddash$magnetic" phase, whose weak coupling regime corresponds to 
the confined SQCD in the $\ddash$electric" phase.  
The anomalies in the $\ddash$electric" theory turn out to be balanced by those from 
hypothetical $\ddash$magnetic" quarks.  
However, in spite of the well-motivated physical view, 
one may consider a conventional option that the theory undergoes successive 
spontaneous symmetry breakdown until the anomaly-matching is realized 
by Nambu-Goldstone superfields together with other chiral composite 
superfields\cite{AnomalyMatch}.  The dynamical requirement on the 
phase transition from the $\ddash$electric" phase to the $\ddash$magnetic" phase 
is replaced by an alternative requirement on the spontaneous chiral 
symmetry breakdown in the $\ddash$electric" phase. Both types of dynamics in SQCD is 
equally possible to be realized.   

In the present article, the dynamics of chiral symmetry breaking is examined to 
gain more insight into physics of SQCD.  Since SUSY is broken in 
the real physics, the smooth SUSY limit of  
softly broken SQCD\cite{Peskin,OtherSoft} is also emphasized in our study.
Our analyses are performed by the use of an effective superpotential 
of the Veneziano-Yankielowicz type\cite{VY,MPRV,RecentVY}, which, in the SUSY 
vacuum, becomes equivalent to Seiberg's superpotential\cite{Seiberg,Legendre}.  
It will then be demonstrated that there is a solution that indicates spontaneous 
chiral symmetry breaking.  Once one vacuum expectation value (VEV) is 
set to be non-vanishing, the dynamics forces other VEV to be generated 
so that at most either $SU(N_f-N_c)$ or $SU(N_f-N_c+1)$ becomes a residual nonabelian 
chiral symmetry. In the SUSY breaking phase, all remaining chiral symmetries 
will be broken.  In $\S$2, our superpotential is formulated. 
In $\S$3, the properties of 
the superpotential together with soft 
SUSY breaking terms are examined. The last section is devoted to a summary.   

\section{\bf Superpotential}
Our general strategy is to utilize an arbitrariness 
appearing in the effective superpotential that cannot be eliminated by a 
symmetry principle only\cite{MPRV,Yasue} just as in Seiberg's superpotential 
for SQCD with $N_f$ = $N_c$+1.  The superfields describing low-energy massless 
spectra are assumed to come from chiral meson superfields ($T$) composed of 
a quark-antiquark pair and chiral (anti)baryon superfields ($B(\bar{B})$) 
by $N_c$-quarks ($N_c$-antiquarks).  Furthermore, chiral exotic meson 
superfields ($U$) composed of ($N_c-1$)-quarks and ($N_c-1$)-antiquarks 
are allowed to participate in our analyses.  
Combinations of color-singlet states such as (with 
abbreviated notations), $BT^{N_f-N_c}\bar{B}/{\rm det}(T)$ (=$Z_B$) and 
$UT^{N_f-N_c+1}/{\rm det}(T)$ (=$Z_U$), are totally neutral under the entire set of chiral 
symmetries as well as an anomalous $U(1)$ symmetry.  
Therefore, an effective superpotential determined by a symmetry principle can 
involve any function of $Z_B$ and $Z_U$ as $f(Z_B, Z_U$).  
The formation of $\langle 0|Z_B |0 \rangle$ ($\neq$ 0) 
(or $\langle 0|Z_U |0 \rangle$) 
leads to the unbroken chiral $SU(N_f-N_c)$ (or $SU(N_f-N_c+1)$) symmetry.  
For $SU(N_f-N_c)$, massless fields are contained in $T$, $B$ and ${\bar B}$ 
while for $SU(N_f-N_c+1)$ they are contained in $U$ as well.

Our effective superpotential, $W_{\rm eff}$, for $N_c$ $>$ 2 is defined as
\begin{equation}\label{Eq:Weff}
W_{\rm eff}=S 
\left\{ 
\ln\left[
\frac{
	S^{N_c-N_f}{\rm det}\left(T\right) f(Z_B, Z_U)
}
{
	\Lambda^{3N_c-N_f}
} 
\right] 
+N_f-N_c\right\},
\end{equation}
where $\Lambda$ is the scale of SQCD and 
$S$ is a color-singlet bilinear in a chiral gauge superfield, whose scalar component 
is $\lambda\lambda$, with $\lambda$ a gaugino.  The decoupling property is manifest, 
as discussed in Ref. 
\cite{VY}. 
Because the 
anomaly-matching conditions are not satisfied in the full chiral symmetry, it must 
spontaneously break down to the next stage where all anomalies present in residual 
chiral symmetries are consistently generated by massless composite superfields.   
 The SUSY vacuum characterized by $\pi_i$ ($i$ = 1$\sim$ $N_f$) = $\pi_b$ = $\pi_{
\bar b}$ = $\pi_u$ = $\pi_\lambda$ = 0 is not dynamically allowed.   
The $\pi$ denote scalar components of superfields defined by
\begin{eqnarray}\label{Eq:ScalarVacuum}
&	& \pi_i=\langle 0|T_i^i |0 \rangle,  \ \pi_b=\langle 0| B^{[12{\dots}N_c]} |0 
\rangle, \ \pi_{\bar b}=\langle 0| {\bar B}_{[12{\dots}N_c]} |0 \rangle,  
 \nonumber \\
&	& \pi_u=\langle 0|U^{[12{\dots}N_c-1]}_{[12{\dots}N_c-1]}|0 \rangle, \ \pi_
\lambda=\langle 0|S|0 \rangle 
\end{eqnarray}
with
\begin{eqnarray}\label{Eq:FiledContent}
  &  &      T_j^i  =  \sum_{A=1}^{N_c} Q_A^i{\bar Q}^A_j, \ 
S = \frac{1}{32\pi^2}\sum_{A,B=1}^{N_c} W_A^BW_B^A,
 \nonumber \\
   &  &    B^{[i_1i_2{\dots}i_{N_c}]} = \sum_{A_1\dots A_{N_c}=1}^{N_c}\frac{1}{N_c!}
                                                 \varepsilon^{A_1A_2{\dots}A_{N_c}}Q_{A_1}^{i_1}\dots 
Q_{A_{N_c}}^{i_{N_c}}, 
 \nonumber \\
    &  &   {\bar B}_{[i_1i_2{\dots}i_{N_c}]} = \sum_{A_1\dots A_{N_c}=1}^{N_c}\frac{1}{N_c!}
                                                 \varepsilon_{A_1A_2{\dots}A_{N_c}}Q_{i_1}^{A_1}\dots 
Q_{i_{N_c}}^{A_{N_c}}, 
 \nonumber \\
    &   &  U_{[j_1j_2{\dots}j_{N_c-1}]}^{[i_1i_2{\dots}i_{N_c-1}]} = \sum_{A=1}^{N_c}C^{A[
i_1i_2{\dots}i_{N_c-1}]}{\bar C}_{A[j_1j_2{\dots}j_{N_c-1}]},
\end{eqnarray}
where $C^{A[i_1i_2{\dots}i_{N_c-1}]}$ =  $\sum_{A_1\dots A_{{N_c}-1}}$
$\varepsilon^{AA_1A_2{\dots}A_{{N_c}-1}}Q_{A_1}^{i_1}\dots Q_{A_{N_c-1}}^{i_{N_c-1}}
/(N_c-1)!$, and the similarly of ${\bar C}$ 
and $Q_A^i$, and ${\bar Q}_i^A$ and $W_A^B$, respectively, represent chiral superfields of quarks, 
antiquarks, and gluons. 
The behavior of $W_{\rm eff}$ in the limit of vanishing gauge coupling $g$ is readily 
found, by applying the rescaling 
$S$ $\rightarrow$ $g^2S$ and invoking the definition $\Lambda$ $\sim$ $\mu\exp (-8\pi^2/(3N_c-N_f)g^2)$, to be $W_{\rm eff}$ $\rightarrow$ $WW/4$, which is the tree 
superpotential for the gauge kinetic term.

To proceed to discussing effects from soft SUSY breakings, let us include breaking 
terms in the potential $V$ = $V_{\rm SUSY}$ + $V_{\rm soft}$.  The SUSY-invariant $V_{\rm SUSY}$ 
is defined by
\begin{eqnarray}\label{Eq:V_SUSY}
V_{\rm SUSY} & = & G_T \bigg( 
                     \sum_{i=1}^{N_f}  |W_{{\rm eff};i}|^2         
                \bigg)
          + G_B\bigg(
			\sum_{i=b,{\bar b}} |W_{{\rm eff};i}|^2 
       	  \bigg)
 \nonumber \\
	   & & {\hskip 30mm} +G_U  |W_{{\rm eff};u}|^2
         +G_S |W_{{\rm eff};\lambda} |^2,
\end{eqnarray}
where $W_{{\rm eff};i}$ =  $\partial W_{\rm eff}/\partial \pi_i$, etc., and  $G_T$ = $G_T(T^{\dagger}T)$ 
characterizes the kinetic term for $T$ defined by the K$\ddot{\rm a}$hlar potential, $K$, 
which is assumed to be diagonal, $\partial^2 K/\partial T_i^{k\ast}\partial T_j^\ell$ 
= $\delta_{ij}\delta_{k\ell}G_T^{-1}$, and similarly for $G_B$ = $G_B(B^{\dagger}B+{\bar B}^\dagger{
\bar B})$, 
$G_U$ = $G_U(U^{\dagger}U)$ and $G_S$ = $G_S(S^{\dagger}S)$.  The SUSY-breaking is 
induced by soft breaking masses denoted by $\mu_{Li}$, $\mu_{Ri}$, $\mu_i$ and $m_{\lambda}$ 
through ${\cal L}_{\rm mass}$ for scalar quarks $\phi_i^A$, scalar antiquarks ${\bar \phi}_i^A$, 
and gluinos $\lambda_A^B$ with Tr($\lambda$)=0 ($A$, $B$=1$\sim$$N_c$) generally expressed as
\begin{equation}\label{Eq:Lmass}
 -{\cal L}_{\rm mass}  =  \sum_{i=1}^{N_f}{\big[}\mu_{Li}^2|\phi_i|^2 + \mu_{Ri}^2|{\bar \phi}_i|^2 
+\mu_i^2(\phi_i {\bar \phi}_i+\phi_i^\ast {\bar \phi}_i^\ast ){\big]} 
           + m_{\lambda}(\lambda\lambda+\bar{\lambda}\bar{\lambda}).
\end{equation}
It is not necessary for subsequent discussion that all of the $\mu$ be effective. 
In contrast to $\mu_{Li, Ri}$, the $\mu_i$ explicitly break chiral $SU(N_f)$ symmetry 
and $m_\lambda$ breaks chiral $U(1)$ symmetry. 
However, once the breaking masses are generated, their mass scales are expected to be 
of the same order.   
Since physics very near the SUSY-invariant vacua  
is our main concern, all breaking masses are 
kept much smaller than $\Lambda$.   For composite superfields, ${\cal L}_{\rm mass}$ is 
translated into $V_{\rm soft}$, which can be cast into the following form 
(with higher orders in scalar masses neglected):  
\begin{eqnarray}\label{Eq:V_soft}
 V_{\rm soft} & = & \left\{ \sum_{i,j=1}^{N_f} \left[ (\mu_{Li}^2 + \mu_{Rj}^2 ) 
                                   \left( \Lambda^{-2}|T_j^i|^2 + \Lambda^{-2(2N_c-3)} |U_j^i|^2 
\right)
                                   +\mu_i^2(T_i^i+T_i^{\dagger i})\right]  \right. 
 \nonumber \\
         &  &  +\left. \Lambda^{-2(N_c-1)} \sum_{i=1}^{N_f}\left( 
                                         \mu_{Li}^2|B^i|^2 + \mu_{Ri}^2|{\bar B}_i|^2
                                   \right) + m_{\lambda}(S+S^\dagger ) \right\} \bigg|_{
\theta={\bar \theta}=0}.
\end{eqnarray}
Here 
$| U_j^i |^2$ $\sim$ $\sum_{i_2\cdots i_{N_c-1}}\sum_{j_2\cdots j_{N_c-1}}|U_{[jj_2\cdots j_{N_c-1}]}^{[ii_2\cdots i_{N_c-1}]}|^2$,  
$| B^i |^2$ $\sim$ $\sum_{i_2\cdots i_{N_c}}|B^{[ii_2\cdots i_{N_c}]}|^2$ 
and similarly for $ | {\bar B}_i |^2$.  

\section{\bf Spontaneous breaking}
Now, let us consider a confining phase where $T$, $B$ and ${\bar B}$ serve as massless 
composites.  It will be shown that the favored number of $\pi_i$ with VEV = ${\cal O}(\Lambda^2 )$ is 
$N_c$.  Since the dynamics requires that some of the $\pi$ acquire non-vanishing 
VEV's, suppose that one of the $\pi_i$ ($i$=1 $\sim$ $N_f$) develops a VEV, and let this be  
labeled by $i$ = $1$: $|\pi_1|$ = $\Lambda_T^2$ $\sim$ $\Lambda^2$.  The 
conditions on $\pi_i$ 
including $\langle 0|Z_B|0 \rangle$ = $z_B$ are simply given by
\begin{equation}
\label{Eq:Simplest_Ta_SUN}
G_TW_{{\rm eff};i}^\ast
\frac{\pi_\lambda}{\pi_i} \left( 1-\alpha_B \right) 
 =  G_SW_{{\rm eff};\lambda}^\ast \left( 1-\alpha_B \right) + M_i^2 +\beta_B X_B
\end{equation}
for $i$=1$\sim N_c$ and by (\ref{Eq:Simplest_Ta_SUN}),  
with $\alpha_B$ = $\beta_B$ = 0, for $i$=$N_c+1\sim N_f$, 
where $\alpha_B$ = $z_Bf^\prime (z_B)/f(z_B)$,  $\beta_B$ = $z_B\alpha_B^\prime$, and  
\begin{eqnarray}
\label{Eq:ScalarMass}
& & M_i^2 = \left( \mu_{Li}^2+\mu_{Ri}^2\right) \bigg| \frac{\pi_i}{\Lambda} \bigg|^2 + 
\mu_i^2\pi_i + G_T^\prime \big| \pi_i\big|^2\sum_{j=1}^{N_f}\big| W_{{\rm eff};j}\big|^2, 
\\
\label{Eq:X_B}
&& X_B = G_T\sum_{i=1}^{N_c}W_{{\rm eff};i}^\ast\frac{\pi_\lambda}{\pi_i} -
G_B\sum_{x=b,{\bar b}}W_{{\rm eff};x}^\ast\frac{\pi_\lambda}{\pi_x}. 
\end{eqnarray}
It should be noted that the effects of field-dependent kinetic terms 
can be regarded as extra sources of soft SUSY breakings as in (\ref{Eq:ScalarMass}).  
Suppose that some VEV other than $\pi_1$ are zero (or much smaller than $\pi_1$).  Then 
$M_1^2$ $\gg $ $M_i^2$ ($i$ $\neq$ 1) because the SUSY breaking soft masses are of 
the same order.  Equation (\ref{Eq:Simplest_Ta_SUN}) yields
\begin{equation}
\bigg| \frac{\pi_i}{\pi_1} \bigg|^2 \sim 1+\frac{M_1^2}{G_SW_{{\rm eff};\lambda}^\ast (1-\alpha
_B)+\beta_BX_B},
\end{equation}
for $i$ = 2 $\sim$ $N_c$, from which $|\pi_i|$ $\rightarrow$  $|\pi_1|$ = $\Lambda_T^2$ 
is derived in the SUSY limit defined by $M_1^2$ $\rightarrow$ 0 as long as 
$M_1^2$ $\sim$ $G_SW_{{\rm eff};\lambda}^\ast (1-\alpha_B)$ + $\beta_BX_B$, which will be the case.  
This behavior implies that, in the SUSY limit, (\ref{Eq:Simplest_Ta_SUN}) forces all 
$\pi_i$ ($i$=1$\sim N_c$) be of the same order;  i.e. 
$G_TW_{{\rm eff};1}^\ast${${\pi_\lambda}/{\pi_1}$} = $\cdots$ = $G_TW_{{\rm eff};N_c}^\ast${${\pi_\lambda}/{\pi_{N_c}}$} 
giving $\pi_i$ $\sim$ $\Lambda^2$, although ${\pi_\lambda}/{\pi_i}$ = 0 is satisfied for any values of 
$\pi_i$ including $\pi_i$ = 0 as long as $\pi_\lambda$ = 0.     
On the other hand, in the extreme case, where $\mu_{L1, R1, 1}$=0, there is a solution 
for which $\pi_{i=2 \sim N_f}$ = $\pi_{b(\bar b)}$ = $\pi_\lambda$ = 0 
in the SUSY limit.  However, this is not dynamically allowed, since the  
anomaly-matching conditions for the residual chiral symmetries 
are not fulfilled. Therefore, in the general case, 
where all soft breaking masses are of the 
same order, the SUSY vacuum is characterized by 
\begin{equation}
|\pi_{i=1\sim N_c}| = \Lambda_T^2, 
\end{equation}
thus yielding $SU(N_c)_{L+R}$.  In other words, once the spontaneous breaking is triggered, then 
$|\pi_{i=1\sim N_c}|$ = $\Lambda_T^2$ is a natural solution of SQCD, where the soft SUSY breakings 
can be consistently introduced. 

From the constraint $W_{{\rm eff};\lambda}$ = 0, for the exact SUSY
\begin{equation}
\label{Eq:fZ_B}
f(z_B) = \prod_{i=N_c+1}^{N_f}\left( \frac{\pi_\lambda}{\Lambda\pi_i}\right) \cdot 
\prod_{a=1}^{N_c}\left( \frac{\Lambda^2}{\pi_a}\right)
\end{equation}
can be derived.  The function $f(z_B)$ should satisfy $f(z_B)$ = 0, because of 
$\pi_\lambda/\pi_{i=N_c+1 \sim N_f}$ = 0 from $W_{{\rm eff};i}$ = 0 and 
$|\pi_{a=1\sim N_c}|$ = $\Lambda_T^2$.  
It is consistent to demand that $f(z_B)$ = 0 provides the classical constraint of  
${\rm det}(T)$ = $BT^{N_f-N_c}{\bar B}$.  The simplest form of $f(Z_B)$ is then taken to be
\begin{equation}
f(Z_B) = \left( 1 -Z_B \right)^\rho \ \ (\rho > 0)
\end{equation}
from which, due to  $z_B$ $\equiv$ $\pi_b\left( \prod_{i=N_c+1}^{N_f}\pi_i\right)\pi_{\bar b}/\prod_{i=1}^{N_f}\pi_i$ = 
$\pi_b\pi_{\bar b}/\prod_{i=1}^{N_c}\pi_i$ = 1,  
\begin{equation}
\label{Eq:VEV_B}
|\pi_b| \sim |\pi_{\bar b}| \sim \Lambda_T^{N_c}
\end{equation}
is derived.  The influence of the SUSY breaking arises 
through $f(z_B)$ as a tiny deviation from zero: 
$f(z_b)$ = $\left( 1-z_b\right)^\rho$ $\equiv$ $\xi^\rho$ for $\xi$ $\ll$ 1.  This 
behavior of $f(z_B)$ allows us to employ $\alpha_B$ $\sim$ $-\rho/\xi$ 
and $\beta_B$ $\sim$ $-\rho/\xi^2$, whose magnitude is much larger than unity.  

With this in mind, we further calculate the constraints (\ref{Eq:Simplest_Ta_SUN}) by inserting 
$W_{{\rm eff};i=1\sim N_c}$ = $(1-\alpha_B)\pi_\lambda/\pi_i$ and 
similar constraints on $\pi_{i=N_c+1\sim N_f}$ and $\pi_{b({\bar b})}$, which turn out to be 
\begin{eqnarray}
\label{Eq:W_eff_T_N+1-N_f}
& & G_T\bigg| \frac{\pi_\lambda}{\pi_{i=N_c+1 \sim N_f}} \bigg|^2 = 
G_SW_{{\rm eff};\lambda}^\ast + M_i^2,
\\
\label{Eq:W_eff_T_1-N}
& & G_T\bigg| \frac{\rho\pi_\lambda}{\xi\pi_{i = 1 \sim N_c}} \bigg|^2 = 
\frac{1}{N_c-2} 
\left[ 
	(\rho -2)G_SW_{{\rm eff};\lambda}^\ast  - {\cal M}^2
\right]
+ M_i^2,
\\
\label{Eq:W_eff_B}
& & G_{B(\bar B)}\bigg| \frac{\rho\pi_\lambda}{\xi\pi_{b({\bar b})}} \bigg|^2 = 
\frac{1}{N_c-2} 
\left[ 
	(N_c-\rho)G_SW_{{\rm eff};\lambda}^\ast  + {\cal M}^2
\right]
+ M_{b({\bar b})}^2,
\end{eqnarray}
where ${\cal M}^2$ = $\sum_{i=1}^{N_c}M_i^2$ + $\sum_{x=b,{\bar b}}M_x^2$ with 
\begin{equation}
M_x^2 = \sum_{i=1}^{N_c}\mu_{Li(Ri)}^2
\big| \frac{\pi_x}{\Lambda^{N_c-1}} \big|^2 + 
G_B^\prime\big| \pi_x\big|^2\sum_{y=b,{\bar b}} \big|W_{{\rm eff};y}\big|^2.
\end{equation}
The relations (\ref{Eq:W_eff_T_N+1-N_f})-(\ref{Eq:W_eff_B}) show that 
\begin{equation}
\label{Eq:SUSYbreakingVEV}
G_T\big|\pi_{b({\bar b})} \big|^2 \sim G_{B({\bar B})}\big|\pi_{i=1 \sim N_c} \big|^2, \  \ 
\big|\pi_{i=N_c+1 \sim N_f} \big|  \sim 
\xi\big|\pi_{i=1 \sim N_c} \big| ,  
\end{equation}
which are also consistent with (\ref{Eq:VEV_B}).  From the relation (\ref{Eq:fZ_B}), 
$\pi_\lambda$ is calculated to be
\begin{equation}
\big|\pi_\lambda\big| \sim \Lambda^3\xi^{\frac{\rho+N_f-N_c}{N_f-N_c}}.
\end{equation}
These solutions indicate, in softly broken SQCD, the breakdown of 
all chiral symmetries that is in agreement with the result of the dynamics  
of ordinary QCD\cite{QCD}. In the SUSY limit of  
$\pi_{i=N_c+1 \sim N_f}$ $\rightarrow$ 0 owing to
$\xi$ $\rightarrow$ 0, chiral $SU(N_f-N_c)$ symmetry is preserved.  
The constraint on the form of the $W_{\rm eff}$ will arise to 
ensure the positivity of the right-hand side of (\ref{Eq:W_eff_T_N+1-N_f}) 
- (\ref{Eq:W_eff_B}).   For instance,  (\ref{Eq:W_eff_T_1-N}) and  
(\ref{Eq:W_eff_B}) using the definition of ${\cal M}^2$ give $W_{{\rm eff};\lambda}^\ast$ $>$ 0.  
It should be noted that $M_b^2$ can take any values, 
depending upon the explicit form of $G_B$, 
because the term $G_B^\prime\big| \pi_b\big|^2\big|W_{{\rm eff};b}\big|^2$ 
turns out to be $\sim\mu_{Li(Ri)}^2\Lambda^2$ as implied by (\ref{Eq:W_eff_B}) 
since $W_{{\rm eff};b}$ = $\rho\pi_\lambda /\xi\pi_b$, and similarly 
for $M_{\bar b}^2$ and $M_{i=1\sim N_c}^2$.

For the case in which $U$ as well as $T$, $B$ and ${\bar B}$ are sources of massless 
composites, constraints on $\pi_{i = 1\sim N_c-1}$ are given by
\begin{eqnarray}
 G_TW_{{\rm eff};i}^\ast \frac{\pi_\lambda}{\pi_i} \left( 1-\alpha_B -\alpha_U \right) 
 & = & G_SW_{{\rm eff};\lambda}^\ast \left( 1-\alpha_B -\alpha_U \right) + M_i^2
\nonumber \\
\label{Eq:Simplest_Ta_SUN-1}
 & + & \left( \beta_B X_B + \beta_U X_U \right),
\end{eqnarray}
where $\alpha_U$ = $z_Uf_U^\prime (z_U)/f_U(z_U)$,  
$\beta_U$ = $z_U\alpha_U^\prime$ and   
\begin{equation}\label{Eq:X_U}
X_U = \sum_{i=1}^{N_c-1}G_TW_{{\rm eff};i}^\ast\frac{\pi_\lambda}{\pi_i}-
G_UW_{{\rm eff};u}^\ast\frac{\pi_\lambda}{\pi_u}.
\end{equation}
Here, $f(Z_B, Z_U)$ = $f_B(Z_B)f_U( Z_U)$ is assumed for simplicity.  It can be proved that the relation
\begin{equation}\label{Eq:SU_M-N+1}
 f_B(Z_B) =\exp (Z_B), 
\end{equation}
giving $Z_B$ in $W_{\rm eff}$, ensures $\pi_b$ = $\pi_{\bar b}$ = 0, even in 
the SUSY-breaking phase.  This form of $f_B$ simply gives 
$\alpha_B$ = $\beta_B$ = $z_B$ = 0 and further reduces (\ref{Eq:Simplest_Ta_SUN-1}) to
\begin{equation}
\label{Eq:Simplest_Ta_SUN-1_0}
G_TW_{{\rm eff};i}^\ast
\frac{\pi_\lambda}{\pi_i} \left( 1-\alpha_U \right) 
 =  G_SW_{{\rm eff};\lambda}^\ast \left( 1-\alpha_U \right) + M_i^2 + \beta_U X_U.
\end{equation}
By the same reasoning as in the previous case, one concludes that 
\begin{equation}
|\pi_{i=1\sim N_c-1}| = \Lambda_T^2,  \ \ |\pi_u| = \Lambda_T^{N_c-1}, \ \ 
|\pi_{i=N_c \sim N_f}| = \xi\Lambda_T^2, \ \
|\pi_\lambda | \sim \Lambda^3\xi^{\frac{\rho+N_f-N_c+1}{N_f-N_c}},
\end{equation}
which indicate that $SU(N_c-1)_{L+R}$ $\times$ $SU(N_f-N_c+1)_L$ $\times$ $SU(N_f-N_c+1)_R$
remains unbroken at $\xi$ = 0.  The arbitrary function, $f(Z_B, Z_U)$, is 
described by
\begin{equation}
f(Z_B, Z_U) = \exp(Z_B)\left( 1 -Z_U \right)^\rho,
\end{equation}
yielding the classical constraint of ${\rm det}(T)$ = $UT^{N_f-N_c+1}$.  

\section{\bf Summary}
Summarizing our discussions, we have shown that the dynamical 
symmetry breaking of SQCD with $N_f$ $\geq$ $N_c+2$ ($N_c$ $>$ 2) 
in the $\ddash$electric" phase can be handled by either
\begin{equation}
W_{\rm eff}=S 
\left\{ 
\ln\left[
\frac{
	S^{N_c-N_f}{\rm det}\left(T\right) 
	\left(1- \frac{BT^{N_f-N_c}\bar{B}}{{\rm det}(T)}\right)^\rho
}
{
	\Lambda^{3N_c-N_f}
} 
\right] 
+N_f-N_c\right\},
\end{equation}
or 
\begin{equation}
W_{\rm eff}=S 
\left\{ 
\ln\left[
\frac{
	S^{N_c-N_f}{\rm det}\left(T\right) 
	\left(1- \frac{UT^{N_f-N_c+1}}{{\rm det}(T)}\right)^\rho
}
{
	\Lambda^{3N_c-N_f}
} 
\right] 
+ \frac{BT^{N_f-N_c}\bar{B}}{{\rm det}(T)}
+N_f-N_c\right\},
\end{equation}
for $\rho$ $>$ 0.  Seiberg's superpotential for SQCD with $N_f$ = $N_c$+1 that 
corresponds to the $\rho$=1 case is 
not unique in the sense that the former case with determined parameter, $\rho$, 
also describes the same physical properties including the decoupling property. 
Our main finding is that $W_{\rm eff}$ in the present form 
dynamically triggers the successive formation of condensates once one 
VEV such as $\langle 0|T_1^1 |0 \rangle \big|_{\theta=0}$ 
is made non-vanishing.  Such successive formation can be made 
visible by watching the behavior of SQCD with soft SUSY breakings in its 
SUSY limit.  It has been demonstrated that, to be consistent, 
soft SUSY breakings are constrained to include terms of  
the scalar components with non-vanishing VEV\cite{Note}.  This is reasonable since 
$M_i^2$ ($i$=fields with VEV ($\sim$ $\Lambda$)) $\gg$ $M_i^2$ 
($i$=fields without VEV) as long as the SUSY-breaking masses 
are of the same order.  In the SUSY limit, the residual symmetry turns out to include either 
$SU(N_c)_{L+R}$ $\times$ $SU(N_f-N_c)_L$ $\times$ $SU(N_f-N_c)_R$  
or $SU(N_c-1)_{L+R}$ $\times$ $SU(N_f-N_c+1)_L$ $\times$ $SU(N_f-N_c+1)_R$. 
The SUSY breaking further induces spontaneous breakdown of the residual nonabelian 
chiral symmetry as in (\ref{Eq:SUSYbreakingVEV}), 
which is in accord with the result in ordinary QCD physics\cite{QCD} that all 
chiral symmetries are spontaneously broken.  The details of the anomaly-matching conditions 
as well as the possible application to physics of composite 
quarks and leptons have been discussed in Ref. 
\cite{Yasue}.  

It should be noted that the present breakings include a spontaneous breakdown 
of vector symmetries such as $SU(N_f)_{L+R}$ to $SU(N_c)_{L+R}$ $\times$ 
$SU(N_f-N_c)_{L+R}$\cite{VectorBreaking}.  A similar breakdown of a vector symmetry has already been 
found to occur in SQCD with $N_f$ = $N_c$, which permits the breaking of 
$U(1)$ of the baryon number. These breakings are precisely determined by 
the dynamics regulated by a relevant effective superpotential, where the anomaly - 
matching is a dynamical consequence. 

Our suggested physical view on symmetry breaking is that the chiral symmetry is 
spontaneously broken to the vectorial $SU(N_c)$ symmetry
for $N_f$ $\leq$ $N_c$ \cite{Seiberg,VY} and  $N_f$ $\geq$ $N_c+2$ 
(or  to  $SU(N_c-1)$ for $N_f$ $\geq$ $N_c+2$) 
as a remnant of QCD.  Perhaps the same breaking is dynamically 
generated in SQCD with $N_f$ = $N_c+1$ although algebraic  anomaly-matching 
consistency allows the solution of the unbroken chiral $SU(N_f)$ symmetry 
in the $\ddash$electric" phase.   
We expect that this physics in the $\ddash$electric" phase 
probably persists in SQCD with $N_f$ $\leq$ 3$N_c$/2 
while SQCD with 3$N_c$/2 $<$ $N_f$ $<$ 3$N_c$ will be in the 
interacting nonabelian Coulomb phase\cite{Seiberg}, 
where the $N$=2 duality can be transmitted.    
Whether $\langle 0|T_i^i |0 \rangle \big|_{\theta=0}$ is vanishing or 
non-vanishing is a purely dynamical problem since both cases give the correct SUSY vacua of 
the different phase.  It is our hope that this issue can be addressed through future 
lattice calculations that reveal the $\ddash$real" physics of SQCD\cite{Lattice}, although 
the possibility of testing our proposed superpotential remains quite remote.

\vspace{3mm}

\noindent
{\it Note Added}:   
There is a work done by N. Arkani-Hamed and R. Rattazzi\cite{AddedRef} who have 
discussed the similar subject on an instability in the $\ddash$magnetic" phase of SQCD 
near the SUSY limit and also have pointed out the possibility of the spontaneous breaking 
of $SU(N_f)_{L+R}$. 

\section*{\bf Acknowledgements}
We are grateful to S. Horata for fruitful discussions and E. Sekiguchi for enjoyable 
conversations.


\begin{references}
\bibitem{Seiberg} N. Seiberg, Phys. Rev. D {\bf 49} (1994), 6857; 
Nucl. Phys. {\bf B435} (1995), 129.
%
\bibitem{EearlySeiberg} I. Affleck, M. Dine and N. Seiberg, Phys. Rev. Lett. {\bf 51} (1983), 1026; 
Nucl. Phys. {\bf B241} (1984), 493; {\bf B256} (1985), 557.
%
\bibitem{tHooft} G. 't Hooft, in {\em Recent Development in Gauge Theories}, 
Proceedings of the Cargese Summer Institute, Cargese, France, 1979, edited 
by G. 't Hooft  {\em et al.}, 
NATO Advanced Study Institute Series B: Physics Vol. 59 (Plenum Press, New York, 1980).
%
\bibitem{SeibergWitten} N. Seiberg and E. Witten, Nucl. Phys. {\bf B426} (1994), 19; {\bf B431} (1994), 484. 
%
\bibitem{Review} K. Intriligator and N. Seiberg, Nucl. Phys. Proc. Suppl. {\bf 45BC} (1996), 1; 
M.E. Peskin, hep-th/9702094 in {\em Proceedings of the 1996 TASI, Boulder, Colorado, USA, 1996}; 
M. Shifman, Prog. Part. Nucl. Phys. {\bf 39} (1997), 1. 
%
\bibitem{CoulombPhase} K. Intriligator and N. Seiberg, Nucl. Phys. {\bf B431} (1994), 551;
P.C. Argyres, M.R. Plesser and A.D. Shapere, Phys. Rev. Lett. {\bf 75} (1995), 1699;
A. Hanany and Y. Oz, Nucl. Phys. {\bf B452} (1995), 283.
%
\bibitem{BrokenN_2} R.G. Leigh and M.J. Strassler, Nucl. Phys. {\bf B447} (1995), 95;
P.C. Argyres, M.R. Plesser and N. Seiberg, Nucl. Phys. {\bf B471} (1996), 159;
M.J. Strassler, Prog. Theor. Phys. Suppl. {\bf No.123} (1996), 373;
N.Evans, S.D.H. Hsu, M. Schwetz and S.B. Selipsky, Nucl. Phys. Proc. Suppl {\bf 52A} (1997), 223;
P.C. Argyres, Nucl. Phys. Proc. Suppl {\bf 61A} (1998), 149;
T. Hirayama, N. Maekawa and S. Sugimoto, Prog. Theor. Phys. {\bf 99} (1998), 843.
%
\bibitem{GapEquation} 
As a preliminary result, see T. Appelquist, A. Nyffeler and S.B. Selipsky, hep-th/9709177 (Sep., 1997), 
although there is a question on the validity of their gap equations.
%
\bibitem{AnomalyMatch} 
T. Banks, I. Frishman, A. Shwimmer and S. Yankielowicz, Nucl. Phys. {\bf B177} (1981), 157.
%
\bibitem{Peskin} O. Aharony, J. Sonnenschein, M.E. Peskin and S. Yankielowicz, Phys. Rev. D 
{\bf 52} (1995), 6157.
%
\bibitem{OtherSoft} E. D'Hoker, Y. Mimura and N. Sakai, Phys. Rev. D {\bf 54} (1996), 7724; 
N. Evans, S.D.H. Hsu and M. Schwets, Phys. Lett. B {\bf 404} (1997), 77;
H.C. Cheng and Y. Shadmi, Nucl. Phys. {\bf B531} (1998), 125; 
S.P. Martin and J. Wells. Phys. Rev. D {\bf 58} (1998), 115013; 
M. Chaichian, W.-F. Chen and T. Kobayashi, Phys. Lett. B {\bf 432} (1998), 120.
%
\bibitem{VY} G. Veneziano and S. Yankielowicz, Phys. Lett. {\bf 113B} (1983), 321; 
T. Taylor, G. Veneziano and S. Yankielowicz, Nucl. Phys. {\bf B218} (1983), 493. 
%
\bibitem{MPRV} A. Masiero, R. Pettorino, M. Roncadelli and G. Veneziano, Nucl. Phys. 
{\bf B261} (1985), 633. 
%
\bibitem{RecentVY} For recent studies, M. Schlitz and M. Zabzine, hep-th/9710125 (Oct., 
1997); 
G.R. Farra, G. Gabadadze and M. Schwetz, Phys. Rev. D {\bf 58} (1998), 015009. 
%
\bibitem{Legendre} K. Intriligator, R. Leigh and N. Seiberg, Phys. Rev D {\bf 50} (1994), 1092.
%
\bibitem{Yasue} M. Yasu${\grave {\rm e}}$, Phys. Rev. D {\bf 35} (1987), 355 and D 
{\bf 36} (1987), 932; 
Prog Theor. Phys. {\bf 78} (1987), 1437.
%
\bibitem{QCD} D. Weingarten, Phys. Rev. D 
{\bf 51} (1983), 1830; C. Vafa and E. Witten, Nucl. Phys. {\bf B234} (1984), 173.
%
\bibitem{Note} This possibility has been overlooked in the previous analyses 
in Ref.
\cite{Yasue}.
%
\bibitem{VectorBreaking} For $SU(N_f)$ $\rightarrow$ $SU(N_c)$ $\times$ $SU(N_f-N_c)$ 
as a color symmetry breaking, see, K. Fujikawa, Prog. Theor. Phys. {\bf 101} (1999), 161.
%
\bibitem{Lattice} For a recent review, I. Montvay, Nucl. Phys. Proc. Suppl. {\bf 63} (1998), 108
and references therein.
%
\bibitem{AddedRef}  N. Arkani-Hamed and R. Rattazzi,  hep-th/9804068 (Apr., 1998); 
\end{references}
\end{document}